\documentstyle[aps,preprint]{revtex}
\def\be{\begin{equation}}
\def\ee{\end{equation}}
\def\bea{\begin{eqnarray}}
\def\eea{\end{eqnarray}}
\begin{document}
\begin{center}
{\LARGE \bf How short can the hair of a black hole be?}
\vskip1cm
{\bf Dar\'\i o N\'u\~nez, Hernando Quevedo, and Daniel Sudarsky}\\
\vskip5mm

Instituto de Ciencias Nucleares\\
Universidad Nacional Aut\'onoma de M\'exico \\
A. P. 70--543, M\'exico, D. F. 04510,  M\'exico. \\
\end{center}

\begin{abstract}
{We show that in all theories in which black hole hair
has been discovered, the region with non-trivial
structure of the non-linear matter fields must extend
beyond $3/ 2$ the horizon radius, independently
of all other parameters present in the theory. We argue
that this is a universal lower bound that applies in
every theory where hair is present. This {\it no
short hair conjecture} is then put forward as a more modest alternative
to the now debunked {\it no hair conjecture}.}
\end{abstract}

\vfill\eject

The no--hair conjecture (NHC) proposed in the early 70's,
(for a detailed review, see \cite{Car}),
was explicitly proven for the Einstein--Maxwell theory
\cite{Proof} and for the Einstein scalar theory \cite{Bek1,Su},
but the idea of its universal validity
was based on a physical argument which suggested that
all  matter fields
present in a black hole space time would eventually be either radiated to
infinity or ``sucked'' into the black hole, except when those fields
were associated with conserved charges defined at asymptotic infinity.

At this point, it seems convenient to define what we call ``hair":
We will say that in a given theory there is black hole hair
when the space time metric  and the configuration of the other fields of
a stationary black hole solution are not completely specified by
the conserved charges defined at asymptotic infinity. Thus, in the language
of Ref. \cite{Pres}, we do not consider secondary hair.

Recently, we have witness the discovery of precisely this type of
black hole hair in multiple theories in which  Einstein's gravity is coupled
with self interacting matter fields. This has been achieved using
numerical analysis in theories like Einstein--Yang--Mills (EYM) \cite{eym},
Einstein--Skyrme (ES) \cite{es}, Einstein--Yang--Mills--Dilaton
(EYMD) \cite{eymd} and Einstein--Yang--Mills--Higgs (EYMH) \cite{eymh}.

These results, which demonstrate the falsehood of the NHC, naturally give
raise the question: What happened to the physical arguments put forward to
support it? It seems clear now that the non-linear character of the matter
content of the examples discussed plays an essential role:
The interaction between the part of the field that would be radiated away
and that which would be sucked in is responsible for the failure of the
argument and, thus, for the existence of black hole hair.

On the other hand, this suggests that the non-linear behavior of the matter
fields must be present both, in a region very close to the horizon
(a region from which presumably the fields  would tend to be sucked in)
and in a region relatively distant from the horizon (a region from which
presumably the fields  would tend to be radiated away), with the
self interaction being responsible for binding together the fields
in these two regions. In this letter, we will present analytic evidence
for this heuristic argument by showing the existence of a lower bound for the
size of the region where the non-linear behavior of the field is present.
This lower bound value is valid for all theories in which black hole hair
has been found, and it is
independent of all parameters present in the theory; more specifically we
will show that in all such cases the region of interest must extend beyond
$3/2$
the horizon radius.

The theories under consideration are described by the Lagrangian
\be
{\cal L}={\cal L}_G + {\cal L}_M \ ,
\ee
where ${\cal L}_G = (16\pi)^{-1} \sqrt{- g}  R $ is the standard Hilbert
Lagrangian for Einstein's gravity, and ${\cal L}_M$ is the Lagrangian for
some set of non-linear matter fields minimally coupled to gravity.
We will use  units in which $c=G=1$.
We will focus on static spherically symmetric gravitational fields and so
we write the line element of the black hole space-time as
\be
ds^2 = -e^{-2\,\delta}\mu\,dt^2 +\mu^{-1}\,dr^2 +r^2\,d\Omega^2 \ ,
\label{eq:lel}
\ee
where $\delta$ and  $\mu=1-2m(r)/r$ are functions of $r$ only, and we
assume that there
is a regular event horizon at $r_H$, so $m(r_H)=r_H/2$.
The matter fields will also respect the symmetries of the space-time and in
particular we will look at the specific ansatz employed in each of the
cases where black hole hair has been discovered.

In all cases considered here, Einstein's equations for the
metric (\ref{eq:lel}) and for the corresponding ansatz for the matter
fields can be
written as:
\be
\delta^\prime  =  \beta\,{\cal K}\ , \qquad
\mu^\prime    =  {1\over r}[ 1-\mu  +\alpha \,({\cal K} + U)],
\label{eq:ee}
\ee
where $\alpha, \beta,\, {\cal K}$ and $U$ take particular values in each case.

The crucial step in showing the validity of our conjecture in
different cases lies in the analysis of the matter field equation
applying the Lyapunov method
in the fashion  recently employed by one of us to prove a no--hair
theorem in Einstein--Higgs theory \cite{Su}. That way of using the
Lyapunov method
allows us to obtain from the matter field equations and Einstein's
equations a generic relationship of the form
$ {\cal E}' = B(r)$ where ${\cal E} \propto e^{-\delta} ({\cal K} - U) $
is a function of
$r$ which is {\it negative} on
the horizon and must tend to zero (or a positive value) at infinity if
we demand the existence of a black hole solution. Hence the function $B(r)$
must be {\it positive} in some region. Furthermore we show that in all
such cases this
region corresponds to values of $r$ that are larger than $3/2$ of the
horizon radius.

We now proceed to prove the conjecture stated above for all cases in
which black hole hair has been found.

$i$) In the pure EYM case, (SU(2)), the matter Lagrangian has the form
\be
{\cal L}_M=-{\sqrt{-g}\over 16\pi f^2}
 {F_{\mu\,\nu}}^a\,{F^{\mu\,\nu}}_a,
\ee
where ${F_{\mu\,\nu}}^a=\partial_\mu {A_\nu}^a - \partial_\nu {A_\mu}^a +
{\epsilon^a}_{bc}{A_\mu}^b\,{A_\nu}^c$, is the field strength for the
gauge field ${A_\mu}^a$, and $f$ represents the gauge
coupling constant. Furthermore, we use the static spherically
symmetric ansatz for the potential
\be
A=\sigma_a\,{A_\mu}^a\,dx^\mu=\sigma_1 \,w\,d\,\theta
+ (\sigma_3\,\cot\theta +
\sigma_2\,w)\sin\theta\,d\phi,
\ee
where $w$ is a function of $r$ only. Einstein's equations for this case
may be written as in Eq.(\ref{eq:ee}),
with ${\cal K} = \mu \,w'^2$, $U = (1-w^2)^2/
(2r^2)$, $\alpha=  -2/f^2$, and $  \beta =-2/(f^2 \mu r)$. Applying
the Lyapunov method, as in \cite{Su}, we find
\be
{\cal E}' \equiv [ r^2 e^{-\delta} ({\cal K} - U)]' = r e^{-\delta} (3\mu -1)
w'^2 \ .
\label{eq:eym}
\ee

{}From the expressions for ${\cal K}$ and $ U$ we see that ${\cal E}$ is
negative at the
horizon because  ${\cal K}(r_H) =0$, (since $\mu(r_H) = 0$),  and $U(r_H) > 0$.
On the other hand, it can be
shown that asymptotic flatness implies that  ${\cal E} \rightarrow 0$ as
$ r\rightarrow \infty$. Accordingly, ${\cal E}$ must be an increasing function
of $r$ in some intermediate region.
It follows then that the right hand side of
Eq.~(\ref{eq:eym})
must become  positive at some point, i.e. we must have ${\cal K}$
substantially different from zero while $3\mu >1$, a condition
which is equivalent to
$r > 3\,m(r)$, and since $m(r)$ is an increasing function, then
we find that this occurs at some point with $r>3\,m(r_H)=3r_H/2$.
This proves our conjecture.

$ii$) In the ES case, the matter Lagrangian is \cite{es}
\be
{\cal L}_M=\sqrt{-g} {f^2\over 4}\,
Tr( \nabla_\mu W \nabla^\mu W^{-1}) +
{\sqrt{-g} \over 32\,e^2}
Tr[ (\nabla_\mu W)\,W^{-1}, (\nabla_\nu )\,W^{-1}]^2,
\ee
where $\nabla_\mu$ is the covariant derivative, $W$ is the $SU(2)$
chiral field, and $f^2$ and $ e^2$ are the coupling constants.
For the $SU(2)$ chiral field we use the hedgehog ansatz
$ W(r) = exp({\bf{\sigma \cdot r}} F(r)) $ where ${\bf \sigma}$ are
the Pauli matrices and ${\bf r}$ is a unit radial vector.

In writing the Einstein equations as in Eq.(\ref{eq:ee}),
we follow Ref. \cite{es} and use the variables
$\tilde r = e\,f\,r$, and $\tilde m (\tilde r) =
efm(r)$ so that the function $\mu$ defined above remains invariant.
Dropping the tilde, the resulting equations are equivalent to Eq.(\ref{eq:ee})
with
${\cal K}= \mu\,(r^2/2 + \sin^2 F(r))F'^2$,
$U = \sin^2F\,[1+\sin^2F\, /(2\,r^2)]$,
$\alpha = -8\pi f^2$, and $\beta =-8\pi f^2/(\mu r)$. Using the
Lyaponov method as before, we obtain
\be
{\cal E}' \equiv  [ e^{-\delta} ({\cal K} - U)]' =
- e^{-\delta} \left[ r\mu F'^2 + { 1-\mu \over r\mu}{\cal K}
- 2 r\left( 1 + {U\over r^2} - \sqrt{ 1 + 2{U\over r^2}}\right)\right].
\label{eq:es}
\ee
As before, the r.h.s. of this equation must be positive in some
region. Moreover, ${\cal E}(r_H)<0$ and, since the asymptotic behavior
of the field equations implies $F(r) \approx 1/r^2$
at infinity, it follows that ${\cal E}\rightarrow 0$ from above. Therefore
there must be a point where ${\cal E}=0$,
i.e. ${\cal K} = U$, and ${\cal E}'>0$.
Considering the r.h.s. of Eq.(\ref{eq:es}) at this point,  we find
\be
- r\mu F'^2 -2r \left ( \sqrt{ 1 + 2{{\cal K}\over r^2}} - 1 \right)
+ {{\cal K}\over r\mu} (3\mu -1) >0 .
\ee
Since the first and second terms of the last equation are negative, we
conclude that $ 3\mu >1$ at this point. This is the same condition as
in case $i)$.

$iii$) We will generalize the EYMD case by adding an arbitrary
(positive semi-definite) potential term $V(\phi)$ (which is
expected to arise in super-strings inspired models \cite{Hor2}). The
corresponding matter Lagrangian is given by \cite{eymd}
\be
{\cal L}_M={\sqrt{-g}\over 4\pi}\left(
{1\over 2}\nabla_\mu \phi \nabla^\mu \phi
-{1\over 4 f^2} e^{2\gamma\phi}  {F_{\mu\,\nu}}^a\,{F^{\mu\,\nu}}_a
- V(\phi) \right) ,
\ee
where $f$ is the gauge coupling constant, $\gamma$ is the
dimensionless dilatonic coupling constant, and  ${F_{\mu\nu}}^a$
is the SU(2) Yang--Mills
field strength. The ansatz for the gauge field configuration is
the same as that given in case $i$), and $\phi=\phi(r)$.

The corresponding Einstein equations can be written in the generic form
(\ref{eq:ee})
with
${\cal K} = {\cal K}_1 + {\cal K}_2$, where
${\cal K}_1= \mu\, \exp(2\gamma\phi) w'^2/f^2$,
${\cal K}_2=\mu r^2 \phi'^2/2$ and
$U = r^2 V(\phi) + \exp(2\gamma\phi) (1-w^2)^2/(2f^2 r^2)$,
$\alpha = -2$, and $\beta=-2/(\mu r)$. Following the same
procedure, which in this case involves the two matter field equations,
we find
\be
{\cal E}' \equiv [ r^2 e^{-\delta} ({\cal K} - U)]' =
r e^{-\delta}\left[-2\,{\cal K}_2 - 4 r^2 V(\phi)
+ (3\mu -1)\, {{\cal K}\over \mu} \right] \label{eq:eyd} .
\ee
Since the first and second terms of the r.h.s. of Eq.(\ref{eq:eyd}) are
negative,
we again find the condition $3\mu > 1$ in order to obtain asymptotically
flat solutions.

$iv$) In the  EYMH case, the matter Lagrangian is given by \cite{eymh}
\be
{\cal L}_M=-{\sqrt{-g}\over 4\pi} \,\left[
{1\over 4f^2} {F_{\mu\,\nu}}^a\,{F^{\mu\,\nu}}_a
+ (D_\mu \Phi)^\dagger\,(D^\mu \Phi) + V(\Phi)\right],
\ee
where $D_\mu $ is the usual gauge--covariant derivative,
$\Phi$ is a complex doublet Higgs field, and ${F^{\mu\,\nu}}_a$ is the
$SU(2)$ Yang--Mills
field given above. In this case, the ansatz for the
Yang--Mills field is the same as before, and for the Higgs field we
have
\be
\Phi={1\over\sqrt{2}}\left(\matrix{0 \cr \varphi (r) \cr}\right).
\ee
The Einstein equations are equivalent to Eq.(\ref{eq:ee})
with
${\cal K} = {\cal K}_1 + {\cal K}_2$, where
${\cal K}_1=\mu r^2 \varphi'^2/2 $, and
${\cal K}_2= \mu  w'^2/f^2$,
$U = r^2 V(\varphi) + (1-w^2)^2/(2f^2 r^2) + (1+w)^2\varphi^2/4$,
$\alpha =- 2$, and $\beta=-2/(\mu r)$. Finally, our procedure
leads to
\be
{\cal E}' \equiv [ r^2 e^{-\delta} ({\cal K} - U)]' =
r e^{-\delta}\left[-2\,{\cal K}_2
- 4 r^2 V(\varphi)
-{1\over 2} (1+w)^2 \varphi^2
+ (3\mu -1)\,{{\cal K}\over \mu} \right]  .
\ee
As in the previous cases, the required behavior of the function ${\cal E}$
leads to the condition $3\mu >1 $ for the region of interest.

This last case is particularly interesting because, in view of the fact
that both, the scalar field and the Yang--Mills field
become massive
through spontaneous symmetry breaking and the Higgs mechanism
respectively, one could
have naively expected that the fields would fall rapidly (in a distance of
the order of the inverse masses) to their asymptotic behavior (exponentially
decreasing behavior) and therefore, that by choosing
the parameters of the theory
appropriately one could obtain a black hole with hair that is as short as
one desires. The above calculation shows explicitly that this {\bf is not}
what happens and thus we view this case as a strong piece of evidence in
support of our conjecture.

This set of results which have been obtained under the
assumption of spherical symmetry, in part because all
the cases in which hair has been discovered also involve this
simplifying assumption, should, we believe, generalize to the
stationary black hole cases where we expect that the region
of non-linear behavior of the matter fields, which we have dubbed the
"Hairosphere" should also be characterized by the length
$r_{Hair} = 3/2 \sqrt{A \over{4 \pi}}$ where $A$ is the
 horizon area. To be more specific we call the ``Hairosphere''
the region of non-linear behavior of the matter fields, (as oppose
to the region where the behavior of the fields is that which is
found asymptotically).
In all cases presented here, there is a change in the behavior of ${\cal E}$
(which starts always as a negative and decreasing function and needs to
increase towards its asymptotic value)
and we have shown that this change always occurs beyond the
point characterized by $r=3\,m(r)>
3\,r_H/ 2$ and this changes implies that the fields have not achieved
their asymptotic behavior.

In view of the evidence shown here and based on the physical arguments
described at the beginning of this letter,
that suggest the existence of such a universal lower bound,
we are lead to conjecture that for all stationary black holes in theories
in which the matter content is described via a Lagrangian in terms of
self interacting fields, the ``Hairosphere'', if it exists,
must extend beyond the above mentioned distance. In short:
{\it If a black hole has hair, then it can not be
shorter that $3/2$ the horizon radius}.

\end{document}